\documentclass[superscriptaddress,twocolumn,secnumarabic,
amssymb,amsmath,nobibnotes,aps,prd,showkeys,showpacs,nofootinbib]{revtex4}%
\usepackage{graphicx}
\usepackage{epsf}
\usepackage{bm}
\usepackage{amsmath}
\usepackage{amsfonts}
\usepackage{amssymb}%
\providecommand{\U}[1]{\protect\rule{.1in}{.1in}}
\newcommand{\be}{\begin{equation}}
\newcommand{\ee}{\end{equation}}

\newcommand{\mincir}{\raise
-3.truept\hbox{\rlap{\hbox{$\sim$}}\raise4.truept\hbox{$<$}\ }}
\newcommand{\magcir}{\raise
-3.truept\hbox{\rlap{\hbox{$\sim$}}\raise4.truept\hbox{$>$}\ }}

\begin{document}
\title{Scalar-Tensor Gravity Cosmology: Noether symmetries and analytical solutions}

\author{Andronikos Paliathanasis}
\affiliation{Faculty of Physics, Department of Astrophysics - Astronomy - Mechanics
University of Athens, Panepistemiopolis, Athens 157 83, Greece}

\author{Michael Tsamparlis}
\affiliation{Faculty of Physics, Department of Astrophysics - Astronomy - Mechanics
University of Athens, Panepistemiopolis, Athens 157 83, Greece}

\author{Spyros Basilakos}
\affiliation{Academy of Athens, Research Center for Astronomy and Applied Mathematics,
Soranou Efesiou 4, 11527, Athens, Greece}

\author{Salvatore Capozziello}
\affiliation{Dipartimento di Fisica, Universita di Napoli "Federico II.}
\affiliation{INFN Sez. di Napoli, Compl. Univ. di Monte S. Angelo, Ed. G., Via Cinthia,
9, I-80126, Napoli, Italy.}
\affiliation{Gran Sasso Science Institute 
(INFN), Viale F. Crispi 7, I-67100, L'Aquila, Italy.}

\begin{abstract}
%In this paper, we present the Noether Symmetry Approach
%in the framework of scalar-tensor cosmology, 
%thus extending the work by Tsamparlis et al. 
%[Gen. Rel. Grav., {\bf 45}, 2004 (2013)].
In this paper, we present a complete Noether Symmetry analysis
in the framework of scalar-tensor cosmology.
Specifically, we consider a non-minimally coupled scalar field 
action embedded in the Friedmann-Lema\^\i tre-Robertson-Walker (FLRW)
spacetime and  provide a full set of Noether
symmetries for related minisuperspaces. The presence 
of symmetries implies that the dynamical system 
becomes integrable and then we can compute cosmological 
analytical solutions for specific functional forms of  coupling and potential
functions selected by the Noether Approach.
\end{abstract}
\date{\today}

\pacs{98.80.-k, 95.35.+d, 95.36.+x}
\keywords{Alternative theories of gravity; Cosmology; conformal transformations; exact solutions}
\maketitle

\hyphenation{tho-rou-ghly in-te-gra-ting e-vol-ving con-si-de-ring
ta-king me-tho-do-lo-gy fi-gu-re}

\section{Introduction}
The discovery of the accelerated expansion of the 
universe 
\cite{Teg04,Spergel07,essence,Kowal08,Hic09,komatsu08,LJC09,BasPli10}
has opened a new path in approaching the cosmological problem. 
Despite the mounting observational evidences on the
existence of the cosmic acceleration, its nature and
fundamental origin is still an open question challenging the very
foundations of theoretical physics.
Usually, the mechanism that is responsible for  cosmic 
acceleration is attributed to new physics which is 
based either on a modified theory of gravity or 
on the existence of some sort of dark energy which is associated
with new fields
in nature (see \cite{Ratra88,curvature,mauro,report,repsergei,
Oze87,Weinberg89,Lambdat,Bas09c,Wetterich:1994bg,
Caldwell98,Brax:1999gp,KAM,fein02,Caldwell,Bento03,chime04,Linder2004,LSS08,
Brookfield:2005td,Boehmer:2007qa,Starobinsky-2007,Ame10} and references therein).

From the mathematical viewpoint, in order to study 
the cosmological features of a particular
\textquotedblleft dark energy\textquotedblright\ model, it is 
essential to specify the
covariant Einstein-Hilbert action of the model and find out the corresponding
energy-momentum tensor. This methodology provides an elegant
way to deal with dark energy in
cosmology. Within this framework, the standard view of the classical scalar field 
dark energy can 
be generalized considering scalar-tensor theories of gravity 
in which the scalar field $\psi$ is 
non-minimally coupled to the Ricci scalar $R$. 
Generally, any theory of gravity that is not simply 
linear in the Ricci scalar can be reduced to a scalar-tensor one, implying 
that among the modified gravity models the scalar-tensor 
theory of gravity is one of the most general case that contains 
also other alternatives (for a review see \cite{Ame10}).  
As an example, the $f(R)$-gravity can be seen as a particular 
case of scalar-tensor gravity obeying the following 
criteria: (a) the scalar field is non-minimally coupled to the Ricci 
scalar and (b) a self-interacting potential is present while there 
is no kinetic term. In this specific case, the scalar field is $\psi =f'(R)$ which is  the first derivative of $f(R)$ function with respect to $R$.
In general, large classes of alternative theories of gravity, non-linear in the curvature invariants or non-minimally coupled in the Jordan frame, can be reduced to general relativity plus scalar field(s) in the Einstein frame \cite{report}.

In a recent paper by the same authors \cite{TT}, 
conformally related metrics and Lagrangians,
in the framework of scalar-tensor cosmology, have been studied. In particular, it has been  proven  that
the field equations of two conformally related Lagrangians are also
conformally related if the corresponding Hamiltonian vanishes.
This is an important feature strictly related to the energy conditions of the theory.
Also, it has been shown that to every non-minimally 
coupled scalar field, we can associate a
unique minimally coupled scalar field in a conformally related space with an
appropriate potential. The existence of such a connection can be used in order
to study the dynamical properties of the various cosmological models, since
the field equations of a non-minimally coupled scalar field can be reduced, at
 conformal level, to  the field equations of the minimally coupled scalar
field. 

With the current work, we complete our previous program
on scalar-tensor theories by calculating the corresponding 
Noether point symmetries as well as the related analytical solutions.
It is interesting to mention that Noether point symmetries have gained a lot of 
attention in cosmology (see \cite{Leach,Cap96, Szy06, Cap07, Capa07, Cap08, Cap09,Vakili08, Yi09, HaoW, CapHam, BB, deSouza, Kucu, Dong}), since they can be 
used as a selection criterion in order to
discriminate the dark energy models, including those of modified gravity 
\cite{BB} as well as to provide analytical solutions. 
Such a program started in \cite{Cap96} where inflationary models have been considered. 
The paradigm can be shortly summarized as follows.
The existence of a Noether symmetry  selects the forms of non-minimal coupling and potential in general scalar-tensor theories of gravity. As a consequence, the related dynamical system which results {\it reduced} because every symmetry is related to a first integral of motion. In most  cases, the presence of such integrals of motion allows to find out general solutions for dynamics.
It is important to stress that by choosing particular classes of metrics, one reduces the field theory to  a point-like one. From a cosmological viewpoint, this means that we are considering dynamical systems defined on {\it minisuperspaces}.  These finite-dimensional dynamical systems are extremely interesting in  Quantum Cosmology (see \cite{CapHam} for a discussion). This consideration is important since allows one to deal with Noether Symmetry Approach 
both in early and late cosmology.

Some remarks are important at this point to relate the Noether Symmetry
Approach to the presence of conserved physical quantities.
Generally speaking, in modified gravitational theories, where 
the Birkhoff theorem is not guaranteed, the
Noether approach can provide a useful tool towards describing 
the global dynamics \cite{Capozziello:2012iea}, through the first integrals 
of motion. Moreover, besides the technical possibility 
of reducing the dynamical system, the
first integrals of motion give always rise to conserved currents that
are not only present in physical space-time but also in configuration
spaces (see the discussion in \cite{TT} and \cite{Cap96}). While 
in space-time such currents are linear momentum, angular momentum etc.
in configuration space the conserved
quantities emerge as relations among dynamical variables, in
particular, among their functions as couplings and self-interaction
potentials. For example, as discussed 
in Capozziello \& Ritis \cite{Rit95}, the presence of Noether symmetries in
scalar-tensor gravity gives rise to an effective cosmological constant and
gravitational asymptotic freedom behaviours induced by potentials and
couplings. This means that, while in the standard spacetime the Noether
charges are directly related to conserved observable quantities, in
the configuration space (minisuperspace), they are present as
''selection rules'' for potentials and coupling functions which are capable of
assigning realistic dynamics. 

In the present work, we complete the 
program started in \cite{Cap96} and \cite{TT}, discussing 
the general structure of scalar-tensor cosmological 
models compatible with the existence of Noether symmetries. 
Moreover, the current work can be seen as a natural continuation 
of our previous works \cite{BB}.
%In the present paper we would like to extend our previous works 
%\cite{TT,BB} by applying the Noether symmetry approach to 
%scalar-tensor theories of gravity. Specifically, 
The layout of the paper is the following. 
In Sec. 2, we present the main ingredients of the dynamical problem
under study. In Secs. 3 and 4 we provide the Noether point symmetries as well 
as the corresponding analytical solutions for the two classes of models considered. We draw our
conclusions in Sec. 5.

\section{The Minisuperspace and the dynamical system}

In the context of scalar-tensor cosmology, let us   
 consider a scalar field $\psi$ (non-minimally) interacting  with the gravitational
field. In this framework, the field equations can be
derived from the following general action 
\begin{equation}
S=\int dt dx^{3}\sqrt{-g}\left[ F\left( \psi,R \right) +\frac{%
\varepsilon }{2}g_{ij}\psi ^{;i}\psi ^{;j}-V\left( \psi \right) %
\right] +S_{m}
\label{EE}
\end{equation}%
where $\varepsilon=\pm 1$, $\psi$ denotes the 
scalar field, $V(\psi)$ is the self-interaction potential, $F(\psi,R)$ is 
the coupling function, $R$ is the Ricci scalar and  
$S_{m}$ is the matter action. The parameter $\varepsilon$ indicates if we are dealing with a regular scalar field or a ghost field.
Assuming a
spatially flat FRW space-time 
\begin{equation*}
ds^{2}=-dt^{2}+a^{2}\left( t\right) \delta _{ij}dx^{i}dx^{j} \;,
\end{equation*}
the infinite degrees of freedom of the field theory reduce to a finite number. In this specific case, the minisuperspace is a 2-dimensional configuration space  defined by the  variables ${\cal Q}=\{\psi,a\}$. The tangent space on which dynamics is defined is ${\cal TQ}=\{\psi,\dot{\psi}, a, \dot{a}\}$ where the dot indicates the derivative with respect to the cosmic time which is the natural affine parameter for  the problem.
Of course, if we consider $F(\psi,R)=R$ then the action (\ref{EE}) boils
down to the nominal, minimally coupled,  scalar field dark energy. 
On the other hand, the $f(R)$ modified gravity is
fully recovered for $F(\psi,R)=f(R)$ and in the absence of 
the kinetic term in the action.
In this study, we consider the case where the coupling function 
is proportional to $R$, $F(\psi,R)=F(\psi)R$. 

Due to the fact that almost every dynamical system is described by a
corresponding Lagrangian, below we apply such ideas 
to the scalar field cosmology. Indeed the 
corresponding Lagrangian and the Hamiltonian (total energy density)
of the field equations are 
\begin{equation}
L=6F\left( \psi \right) a\dot{a}^{2}+6F_{\psi }\left( \psi \right) a^{2}%
\dot{a}\dot{\psi}+\frac{\varepsilon }{2}a^{3}\dot{\psi}^{2}-a^{3}V\left(
\psi \right)  
\label{CLN.15}
\end{equation}
\begin{equation}
E=6F\left( \psi \right) a\dot{a}^{2}+6F_{\psi }\left( \psi \right) a^{2}\dot{%
a}\dot{\psi}+\frac{\varepsilon }{2}a^{3}\dot{\psi}^{2}+a^{3}V\left( \psi
\right) \;. \label{CLN.15e}
\end{equation}%
Note that the Lagrangian (\ref{CLN.15}) is autonomous, hence the
Hamiltonian $E$ is a constant of motion (see also the discussion in \cite{TT}). This constant corresponds to the
trivial Noether point symmetry $\partial _{t}$ (first integral of motion). 

Using the $0 0$ component of the conservation equation  $T_{;\mu}^{\nu\mu}=0~$ we find that the Hamiltonian 
 $E$ is related to the matter
density $\rho _{m}$ as ${\displaystyle \rho _{m}=\frac{\left\vert E\right\vert }{a^{3}}}$. 
%In the case where 
%$E\,$\thinspace $=0$
%then space does not admit dust therefore it has only a scalar field.

Following the technique described in   \cite{TsamGE, TsampJP, BB} 
it is essential to split the Lagrangian (\ref{CLN.15}) 
in the kinematic
part, which defines the kinematic metric (hereafter KM), and the 
remaining part which we
consider to be the potential. Indeed the kinematic metric is written as 
\begin{equation}
ds_{KM}^{2}=12F\left( \psi \right) a\dot{a}^{2}+12F_{\psi }\left( \psi
\right) a^{2}\dot{a}\dot{\psi}+\varepsilon a^{3}\dot{\psi}^{2} \;.
\label{CLN.21}
\end{equation}
The above metric is not the FRW metric of 
the background space-time but a metric defined on the tangent space ${\cal TQ}$. It is related to  the minisuperspace configuration metric in the two 
dimensional space $\{a,\psi \}$. The corresponding
Ricci scalar of the metric $ds_{KM}^{2}$ is computed to be:
\begin{equation}
R_{KM}=\frac{\varepsilon }{4a^{3}}\frac{\left( 2F_{\psi \psi }F-F_{\psi
}^{2}\right) }{\left( \varepsilon F-3F_{\psi }^{2}\right) ^{2}} \;.
\label{CLN.22}
\end{equation} 
Obviously,  knowing $R_{KM}$,  one can estimate $F(\psi)$. 
If we assume that the curvature $R_{KM}$ is constant then  
Eq.(\ref{CLN.22}) implies that $R_{KM}\equiv 0$ 
(due to the presence of $a$ in the denominator) and the minisuperspace is flat. We realize that we need to consider the following two 
cases: (A) the minisuperspace $\{a,\psi \}$ is maximal symmetric (flat $R_{KM}\equiv 0$) 
and (B) the case where the minisuperspace is not necessarily flat but it is 
conformally flat, because all two dimensional spaces are conformally
flat. In the following, we  consider these two situations in detail.

\section{The case of  maximally symmetric $\{a,\psi\}$ minisuperspace}
In this case using the
condition $R_{KM}=0$, Eq.(\ref{CLN.22}) reduces to 
\begin{equation}
2F_{\psi \psi }F-F_{\psi }^{2}=0  \label{CLN.23}
\end{equation}%
and then a solution  is 
\begin{equation}
F\left( \psi \right) =-\frac{F_{0}\varepsilon }{12}\left( \psi +\psi
_{0}\right) ^{2}  \label{CLN.25}
\end{equation}%
where $F_{0}\varepsilon >0$. 

In order to determine the homothetic algebra of the kinematic metric 
(\ref{CLN.21}),  we write it in a more familiar form. 
Actually, in Tsamparlis et al. \cite{TT} we 
introduced the conformal variables $A$, $\Psi$ and ${\cal N}$ by the relations
\begin{equation}
A=\sqrt{-2F}a  \label{CLN.15a} 
\end{equation}
\begin{equation}
d\Psi=\sqrt{\left( \frac{3\varepsilon F_{\psi }^{2}-F}{2F^{2}}\right) }%
~d\psi  \label{CLN.16a}
\end{equation}
\begin{equation}
{\cal N}=\frac{1}{\sqrt{-2F}} \label{CLN.17a} 
\end{equation}
with $F\left( \Psi \right)<0$. 
In the new variables,  
the kinematic metric (\ref{CLN.21}) and the Lagrangian 
(\ref{CLN.15}) 
become
\begin{equation}
ds_{KM}^{2}=N^{2}\left( \Psi \right) \left[ -3A\dot{A}^{2}+\frac{\varepsilon}{2}A^{3}\dot{\Psi}^{2}\right]   
\label{CLN.28}
\end{equation}
\begin{equation}
L=N^{2}\left( \Psi \right) \left[ -3A\dot{A}^{2}+\frac{\varepsilon}{2}%
A^{3}\dot{\Psi}^{2}\right] -A^{3}\bar{V}\left( \Psi \right)  \label{CLN.27}
\end{equation}
where $N^{2}={\cal N}$, $\bar{V}\left( \Psi \right) =N^{6}(\Psi)V\left( \Psi \right)$. 
Also, the coupling function (\ref{CLN.25}) takes the form
\begin{equation}
%F\left( \Psi \right) &=&-\frac{\left\vert F_{0}\right\vert }{12}\exp \left(
%\varepsilon \frac{\sqrt{6}}{3}\sqrt{\frac{\left\vert F_{0}\right\vert }{1+\varepsilon\left\vert
%F_{0}\right\vert }}\Psi \right) \;.\label{CLN.26.6}
%F\left( \Psi \right) =-\frac{\left\vert F_{0}\right\vert }{12}{\rm e}^{\left(
%\varepsilon \sqrt{6}|k|\Psi \right)} 
F\left( \Psi \right) =-\frac{\varepsilon F_{0}}{12}
{\rm e}^{\sqrt{6\varepsilon}|k|\Psi} 
\label{CLN.26.6}
\end{equation}
where 
\begin{equation}
|k|=\frac{1}{3}\sqrt{\frac{|F_{0}|}{|1+\varepsilon F_{0}|}} \;.
\end{equation}
Notice, that the inequality $F_{0}\varepsilon >0$ is satisfied 
either for $\varepsilon=+1$ with $F_{0}>0$ or 
for $\varepsilon=-1$ with $-1<F_{0}<0$. We would like to mention here 
that in the case of $\varepsilon=-1$ with $F_{0}<-1$ one has to replace 
$\Psi$ with $i\Psi$.

We further simplify the above calculations 
by introducing a new coordinate system $(r,\theta)$ defined as
\begin{equation}
r=\sqrt{\frac{8}{3}}A^{\frac{3}{2}}~,~\theta =\sqrt{\frac{3\varepsilon}{%
8}}~\Psi \;. \label{CLN.29}
\end{equation}
Inserting the above variables into Eq.(\ref{CLN.28}), we immediately obtain 
\begin{equation}
ds_{KM}^{2}=N^{2}\left( \theta \right) \left( -dr^{2}+r^{2}d\theta
^{2}\right)   \label{CLN.30}
\end{equation}%
which is directly related to the flat 2D Lorentzian space with metric 
\begin{equation*}
ds^{2}=-dr^{2}+r^{2}d\theta ^{2}
\end{equation*}%
with the conformal factor $N\left( \theta \right)$ 
\begin{equation}
N^{2}(\theta)=N_{0}^{2}{\rm e}^{\mp2|k|\theta} \;\;\;N_{0}^{2}=(\frac{6}{\varepsilon F_{0}})^{1/2} \;.
\label{CLN.300}
\end{equation}
Finally, the Lagrangian takes a rather simple form
\begin{equation}
L=N^{2}\left( \theta \right) \left( -\frac{1}{2}\dot{r}^{2}+\frac{1}{2}r^{2}%
\dot{\theta}^{2}\right) -r^{2}V\left( \theta \right) .  \label{CLN.150a}
\end{equation}%
Armed with the above expressions, we can deduce the homothetic algebra of the
metric from well known previous results 
(see \cite{TsamGE, TsampJP, BB}).

\subsection{Searching for Noether point symmetries}
Let us determine now all the potentials $V\left( \psi \right) $ for
which the above dynamical system admits Noether point symmetries beyond the
trivial one $\partial _{t}$ related to the energy. Subsequently, we shall 
use the resulting 
Noether integrals in order to find out analytical solutions.

For $\left\vert k\right\vert \neq 1$ the homothetic algebra consists of the
gradient  Killing vectors (KVs) 

\begin{eqnarray*}
K^{1} &=&\frac{e^{\left( 1-k\right) \theta }r^{k}}{N_{0}^{2}}\left(
-\partial _{r}+\frac{1}{r}\partial _{\theta }\right) ~,~S_{1}\left( r,\theta
\right) =\frac{r^{1+k}e^{\left( 1+k\right) \theta }}{\left( k+1\right) } \\
K^{2} &=&\frac{e^{-\left( 1+k\right) \theta }r^{-k}}{N_{0}^{2}}\left(
\partial _{r}+\frac{1}{r}\partial _{\theta }\right) ,~S_{2}\left( r,\theta
\right) =\frac{r^{1-k}e^{-\left( 1-k\right) \theta }}{k-1}
\end{eqnarray*}%
(for $\left\vert k\right\vert =1$ see Appendix A) the non-gradient KV 
\begin{equation*}
K^{3}=r\partial _{r}-\frac{1}{k}\partial _{\theta }
\end{equation*}%
and the gradient homothetic vectors (HV)
\begin{equation*}
H^{i}=\frac{1}{N_{0}^{2}\left( k^{2}-1\right) }\left( -r\partial
_{r}+k\partial _{\theta }\right) ~,~H\left( r,\theta \right) =\frac{1}{2}%
\frac{r^{2}e^{2k\theta }}{(k^{2}-1)}.
\end{equation*}%

Specifically, we ask the question: {\it are there 
potentials that can provide non-trivial Noether point 
symmetries and consequently  first integrals of motion}? 
Below we present all possible cases:

\begin{enumerate}
\item First of all, by using the gradient KV $K^{1}$,  we find 

a) for $V\left( \theta \right) =V_{0}e^{2\theta }$,  we have the Noether
symmetries $K^{1},~tK^{1}$ with Noether integrals%
\begin{equation*}
I_{1}=\frac{d}{dt}\left( \frac{r^{1+k}e^{\left( 1+k\right) \theta }}{\left(
k+1\right) }\right) 
\end{equation*}

\begin{equation*}
I_{2}=t\frac{d}{dt}\left( \frac{r^{1+k}e^{\left(
1+k\right) \theta }}{\left( k+1\right) }\right) -\left( \frac{%
r^{1+k}e^{\left( 1+k\right) \theta }}{\left( k+1\right) }\right)
\label{CLN.75a}
\end{equation*}

b) for $V\left( \theta \right) =V_{0}e^{2\theta }-\frac{mN_{0}^{2}}{2\left(
k^{2}-1\right) }e^{2k\theta }$,  we obtain the Noether symmetries $e^{\pm \sqrt{m%
}t}K^{1}$, where $m=$constant, with Noether integrals%
\begin{equation*}
I_{\pm }^{\prime }=e^{\pm \sqrt{m}t}\left[ \frac{d}{dt}\left( \frac{%
r^{1+k}e^{\left( 1+k\right) \theta }}{\left( k+1\right) }\right) \mp \sqrt{m}%
\left( \frac{r^{1+k}e^{\left( 1+k\right) \theta }}{\left( k+1\right) }%
\right) \right]  \label{CLN.99}
\end{equation*}%
From the above 
Noether integrals,  we construct the time independent first integral $%
I_{K^{1}}=I_{+}I_{-}.$

\item The gradient KV $K^{2}$ produces the  Noether symmetries for
the following potentials

a) for $V\left( \theta \right) =V_{0}e^{-2\theta }$,we have the Noether
symmetries $K^{1},~tK^{1}$ with Noether integrals%
\begin{equation*}
J_{1}=\frac{d}{dt}\left( \frac{r^{1-k}e^{-\left( 1-k\right) \theta }}{k-1}%
\right) ~,
%\label{CLN.100}
\end{equation*}
\begin{equation*}
J_{2}=t\frac{d}{dt}\left( \frac{r^{1-k}e^{-\left( 1-k\right)
\theta }}{k-1}\right) -\frac{r^{1-k}e^{-\left( 1-k\right) \theta }}{k-1}
\label{CLN.100}
\end{equation*}

b) for $V\left( \theta \right) =V_{0}e^{-2\theta }-\frac{mN_{0}^{2}}{2\left(
k^{2}-1\right) }e^{2k\theta }$, we have the Noether symmetries $e^{\pm \sqrt{%
m}t}K^{2}$ $m=$constant, with Noether integrals%
\begin{equation*}
J_{\pm }^{^{\prime }}=e^{\pm \sqrt{m}t}\left[ \frac{d}{dt}\left( \frac{%
r^{1-k}e^{-\left( 1-k\right) \theta }}{k-1}\right) \mp \sqrt{m}\frac{%
r^{1-k}e^{-\left( 1-k\right) \theta }}{k-1}\right]  \label{CLN.100.a}
\end{equation*}%
Combining the latter 
Noether integrals,  we construct the time-independent first integral $%
J_{K^{2}}=J_{+}^{\prime }J_{-}^{\prime }.$

\item The non gradient KV $K^{3}$ produces a Noether symmetry for the
potential $V\left( \theta \right) =V_{0}e^{2k\theta }$ with Noether integral 
\begin{equation*}
I_{3}=\frac{re^{2k\theta }}{k}\left( k\dot{r}+r\dot{\theta}\right) .
\label{CLN.100b}
\end{equation*}

\item The gradient HV produces the following Noether symmetries for the
following potentials

a) for $V\left( \theta \right) =V_{0}e^{-2\frac{\left( k^{2}-2\right) }{k}%
\theta }$ , $k^{2}-2\neq 0$ we have the Noether symmetries $2t\partial
_{t}+H^{i}~,~t^{2}\partial _{t}+tH^{i}$ with Noether integrals%
\begin{equation*}
I_{H_{1}}=2tE-\frac{d}{dt}\left( \frac{1}{2}\frac{r^{2}e^{2k\theta }}{(k^{2}-1)}\right) ~,~
\label{CLN.100c}
\end{equation*}%

\begin{equation*}
I_{H_{2}}=t^{2}E-t\frac{d}{dt}\left( \frac{1}{2}\frac{%
r^{2}e^{2k\theta }}{(k^{2}-1)}\right) +\frac{1}{2}\frac{r^{2}e^{2k\theta }}
{(k^{2}-1)} \,. %\label{CLN.100c}
\end{equation*}
We note that in this case the system is the Ermakov-Pinney dynamical system \cite{Ermakov}
and  admits the Noether symmetry algebra  $sl(2,R)$.
%[because it admits the Noether symmetry algebra the $sl(2,R)$ hence the
%Lie symmetry algebra is at least $sl(2,R)$] .

b) For $V\left( \theta \right) =V_{0}e^{-2\frac{\left( k^{2}-2\right) }{k}%
\theta }-\frac{N_{0}^{2}m}{k^{2}-1}e^{2k\theta }~$ , $k^{2}-2\neq 0$ we have
the Noether symmetries $\frac{2}{\sqrt{m}}e^{\pm \sqrt{m}t}\partial _{t}\pm
e^{\pm \sqrt{m}t}H^{i}$ , $m=$constant with Noether integrals%
\begin{equation*}
I_{\pm}=e^{\pm 2\sqrt{m}t}\left[ \frac{1}{\sqrt{m}}E\mp \frac{d}{dt}\left(
\frac{1}{2}\frac{r^{2}e^{2k\theta }}{(k^{2}-1)}\right) +2\sqrt{m}\left( \frac{1
}{2}\frac{r^{2}e^{2k\theta }}{(k^{2}-1)}\right) \right]  \label{CLN.105}
\end{equation*}
This is also the Ermakov-Pinney dynamical system with a linear oscillator.
Therefore it admits the Ermakov - Pinney invariant which we may construct
with the use of the dynamical Noether symmetries %\ref{TsamErm} 
or with the use of the corresponding Killing Tensor.

\item Lastly, the case $V\left( \theta \right) =0$ corresponds to 
the free particle (see \cite{TsamGE}).
\end{enumerate}

\subsection{Analytical solutions}
Using the above Noether symmetries and the 
corresponding integral of motions, we can fully solve 
the dynamical problem of the scalar tensor cosmology.
%%%%%When $\left\vert k\right\vert \neq 1$ we have to consider two cases 
%$k|>1$ and $|k|<1$. 
In order to simplify the analytical solutions, we consider the new variables
\begin{equation}
x=S_{1}(r,\theta)=\frac{r^{1+k}e^{\left( 1+k\right) \theta }}{k+1}~,~y=S_{2}(r,\theta)=\frac{%
r^{1-k}e^{-\left( 1-k\right) \theta }}{k-1}  \label{CLN.97}
\end{equation}
and the inverse transformation is
\begin{eqnarray}
\theta &=&\frac{1}{2|k^{2}-1| }\ln \left[ \frac{|k^{2}-1|^{1-k}}{\left( k-1\right) ^{2}}\frac{x^{1-k}}{y^{1+k}}\right]
\label{CLN.98} \\
r &=&\sqrt{|k^{2}-1| xy}\left[ \frac{|k^{2}-1|^{1-k}%
}{\left( k-1\right) ^{2}}\frac{x^{1-k}}{y^{1+k}}\right] ^{\frac{k}{2\left(
k^{2}-1\right) }} \;. \label{CLN.98a}
\end{eqnarray}
We find that in the new coordinates $(x,y)$,  the
Lagrangian (\ref{CLN.150a})  takes the form%
\begin{equation}
L\left( x,y,\dot{x},\dot{y}\right) =\epsilon_{k}\frac{N_{0}^{2}}{2}\dot{x}\dot{y}%
-U\left( x,y\right)  \label{CLN.115}
\end{equation}%
where $U\left( x,y\right) =r^{2}V\left( \theta \right)$
and $\epsilon_{k}=+1$ for $|k|>1$ ($\epsilon_{k}=-1$ for $|k|<1$).
Note that $V\left(\theta \right)$ are the potentials which have been  
presented in the previous section. 

We would like to stress that 
the solution of the field equations for each potential is a formal and
lengthy operation which adds nothing but unnecessary material to the matter.
What is interesting of course is the final answer for each case and this is
what we show  in a compact presentation below. 
%below for each of the potentials discussed above.
Specifically, the analytical solutions can be categorized into 
seven separate cases  
\begin{itemize}
\item The first class is $U_{1}\left( x,y\right) =V_{0}r^{2}e^{2k\theta }=V_{0}|k^{2}-1|xy$
\begin{eqnarray}
x\left( t\right) &=&x_{1}{\rm Sinn} \left( \omega t\right) +x_{2}{\rm Coss} \left(
\omega t\right)  \label{CLN.123} \\
y\left( t\right) &=&y_{1}{\rm Sinn} \left( \omega t\right) +y_{2}{\rm Coss} \left(
\omega t\right)  \label{CLN.124}
\end{eqnarray}%
where 
\begin{equation}
(\mathrm{Sinn}\omega,\mathrm{Coss}\omega)=\left\{
\begin{array}
[c]{cc}%
(\mathrm{sin}\omega,\mathrm{cos}\omega) & \mbox{$|k|>1$}\\
(\mathrm{sinh}\omega,\mathrm{cosh}\omega) & \mbox{$\;\;\;|k|<1$}
\end{array}
\right.  \label{SSINN}%
\end{equation}
$\omega ^{2}=\frac{2V_{0}|k^{2}-1|}{N_{0}^{2}}$ and the
Hamiltonian is $\ $%
\begin{equation*}
E=V_{0}|k^{2}-1| \left( x_{1}y_{1}+\epsilon_{k}x_{2}y_{2}\right)
\end{equation*}

\item $U_{2}\left( x,y\right) =V_{0}r^{2}e^{2\theta }=V_{0}\left( k+1\right) ^{%
\frac{2}{1+k}}~x^{\frac{^{2}}{1+k}}$,
as long as $k\neq -3$ we have
\begin{eqnarray}
x\left( t\right) &=&x_{1}t+x_{2}  \label{CLN.125} \\
y\left( t\right) &=&-\epsilon_{k}\frac{2\bar{V}\left( k+1\right) \left(
x_{1}t+x_{2}\right) ^{\left( 1+\frac{2}{1+k}\right) }}{x_{1}^{2}\left(
3+k\right) N_{0}^{2}}+y_{1}t+y_{2}\nonumber \\ 
&  & 
.\label{CLN.126}
\end{eqnarray}%
where $\bar{V}=V_{0}\left( k+1\right) ^{\frac{2}{1+k}}$ and the Hamiltonian
is ~%
\begin{equation*}
E=\epsilon_{k}\frac{y_{1}x_{1}N_{0}^{2}}{2}.
\end{equation*}

If $k=-3$ then $y(t)$ becomes
\begin{equation}
y\left( t\right) =-2\frac{\bar{V}}{N_{0}^{2}x_{1}^{2}}\ln \left(
x_{1}t+x_{2}\right) +y_{1}t+y_{2}.  \label{CLN.127}
\end{equation}%
%$x(t)$ $E$\ being the same.

\item 
$U_{3}\left( x,y\right) =V_{0}r^{2}e^{-2\theta }=V_{0}|k-1|^{
\frac{2}{1-k}}y^{\frac{2}{1-k}}$

When $k\neq 3$%
\begin{eqnarray}
x\left( t\right) &=&\frac{2\bar{V}|k-1| \left(
y_{1}t+y_{2}\right) ^{1+\frac{2}{k-1}}}{y_{1}^{2}\left( k-3\right) N_{0}^{2}}%
+x_{1}t+x_{2}  \label{CLN.128} \\
y\left( t\right) &=&y_{1}t+y_{2}  \label{CLN.129}
\end{eqnarray}%
where $\bar{V}=V_{0}|k-1|^{\frac{2}{1-k}}$ 
and the Hamiltonian is 
\begin{equation*}
E=\epsilon_{k}\frac{y_{1}x_{1}N_{0}^{2}}{2}.
\end{equation*}
In this context if $k=3$ then $x(t)$ takes the form  
\begin{equation}
x\left( t\right) =-\frac{2\bar{V}}{N_{0}^{2}y_{1}^{2}}\ln \left(
y_{1}t+y_{2}\right) +x_{1}t+x_{2}.  \label{CLN.130}
\end{equation}%

\item 
$U_{4}\left( x,y\right) =V_{0}r^{2}e^{2\theta }+mr^{2}e^{2k\theta }=\bar{V}_{0}~x^{\frac{^{2}}{1+k}}+\bar{m}xy$,  
%\begin{eqnarray}
%x\left( t\right) &=&x_{1}{\rm Sinn} \left( \omega t+\omega _{0}\right) \\
%y\left( t\right) &=&{\rm Coss} \left( \omega t+\omega _{0}\right) \left( y_{1}+2\epsilon_{K}\frac{\omega }{\bar{m}}\int \frac{E-x_{1}\bar{V}_{0}{\rm Sinn} \left( \omega
%t+\omega _{0}\right) ^{\frac{2}{1+k}}}{x_{1}\left( {\rm Coss} \left( \omega
%t+\omega _{0}\right) +1\right) }dt\right)
%\end{eqnarray}%
in this class we find 
\begin{widetext}
\begin{eqnarray}
x\left( t\right)=x_{1}{\rm Sinn} \left( \omega t+\omega _{0}\right) \\
y\left( t\right)={\rm Coss} \left( \omega t+\omega _{0}\right) \left( y_{1}+2\epsilon_{K}\frac{\omega }{\bar{m}}\int \frac{E-x_{1}\bar{V}_{0}{\rm Sinn} \left( \omega
t+\omega _{0}\right) ^{\frac{2}{1+k}}}{x_{1}\left( {\rm Coss} \left( \omega
t+\omega _{0}\right) +1\right) }dt\right)
\end{eqnarray}
\end{widetext}
where $\bar{V}_{0}=V_{0}\left( k+1\right) ^{\frac{2}{1+k}}~,~
\bar{m}=m|k^{2}-1|$,
$\omega ^{2}=\frac{2\bar{m}}{N_{0}^{2}}$ and $E=y_{2}$.

\item similarly for $U_{5}\left( x,y\right)=r^{2}e^{-2\theta }+mr^{2}e^{2k\theta }=\bar{V}
_{0}y^{\frac{2}{1-k}}+\bar{m}xy$ we obtain
\begin{widetext}
\begin{eqnarray}
x\left( t\right)={\rm Coss} \left( \omega t+\omega _{0}\right) \left( x_{1}+2\epsilon_{k}
\frac{\omega }{\bar{m}}\int \frac{x_{2}-y_{1}\bar{V}_{0}{\rm Sinn} \left( \omega
t+\omega _{0}\right) ^{\frac{2}{1-k}}}{y_{1}\left( {\rm Coss} \left( \omega
t+\omega _{0}\right) +1\right) }dt\right) \\
y\left( t\right) =y_{1}{\rm Sinn} \left( \omega t+\omega _{0}\right)
\end{eqnarray}
\end{widetext}
where $\bar{V}_{0}=V_{0}|k-1|^{\frac{2}{1-k}}$ and $E=x_{2}$.

\item
$U_{6}\left( x,y\right) =V_{0}r^{2}e^{-2\frac{\left( k^{2}-2\right) }{k}%
\theta }+mr^{2}e^{2k\theta }=\bar{V}_{0}\frac{1}{y^{2}}\left( \frac{x}{y}%
\right) ^{\frac{2}{k}-1}+\bar{m}xy$
with $\bar{V}_{0}=V_{0}\frac{|k^{2}-1|^{\frac{2}{k}-1}}{
|k-1|^{\frac{4}{k}}}$. The current dynamical system is the 
so called Ermakov-Pinney system. To solve this dynamical problem, it 
is convenient to go to the following coordinates
%spherical coordinates namely, 
$(x,y)=(ze^{w},ze^{-w})$. 
In this coordinate system we recover the Ermakov-Pinney equation: 
\begin{equation}
\ddot{z}+2\epsilon_{k}\bar{m}z+\epsilon_{k}N_{0}^{2}\frac{J_{EL}}{z^{3}}=0  
\end{equation}%
where $J_{EL}=z^{4}\dot{w}^{2}-2\epsilon_{k}\frac{\bar{V}_{0}}{N_{0}^{2}}e^{\frac{4}{k}w}$ is the Ermakov invariant. The solution of the above differential
equation is
\begin{widetext}
\begin{eqnarray}
z\left( t\right) &=&\left[ l_{0}z_{1}\left( t\right) +l_{1}z_{2}\left(
t\right) +l_{3}\right]^{\frac{1}{2}}  \label{CLN.137} \\
e^{\frac{4}{k}w\left( t\right) } &=&-\epsilon_{k}\frac{N_{0}^{2}J_{EL}}{2\bar{V}_{0}}%
\left[ 1-\tanh ^{2}\left( \frac{2\sqrt{J_{EL}}}{k}\left( \int \frac{dt}{%
z^{2}\left( t\right) }+l_{4}\right) \right)\right]  %\label{CLN.138}
\end{eqnarray}
\end{widetext}
where $z_{1,2}\left( t\right) $ are solutions of the differential equation ~$%
\ddot{z}+2\epsilon_{k}\bar{m}z=0$ and $l_{0-4}$ are constants.

\item Lastly, $U_{7}(x,y)=0$ is the free particle system, 
a solution of which is 
\begin{eqnarray}
x\left( t\right) &=&x_{1}t+x_{2}~,~y\left( t\right) =y_{1}t+y_{2}
\end{eqnarray}
with $E=\epsilon_{k}\frac{N_{0}^{2}}{2}x_{1}y_{1}$.
\end{itemize}

\section{ The case of  2d conformally-flat metric }
In this case the kinetic metric (\ref{CLN.30})
is non-flat (i.e. $R_{KM}\neq
0)$ but, of course, it is conformally flat being a two dimensional metric. Its
conformal algebra is infinity dimensional; however it has a closed subalgebra
consisting of the following vectors (this is the special conformal algebra
of $M^{2}$):%
\begin{eqnarray}
X^{1} &=&\cosh \theta \partial _{r}-\frac{1}{r}\sinh \theta \partial
_{\theta }~~,~X^{2}=\sinh \theta \partial _{r}-\frac{1}{r}\cosh \theta
\partial _{\theta }  \notag \\
X^{3} &=&\partial _{\theta }~~\ ,~X^{4}=r\partial _{r}~~,~X^{5}=\frac{1}{2}%
r^{2}\cosh \theta \partial _{r}+\frac{1}{2}r\sinh \theta \partial _{\theta }
\notag \\
X^{6} &=&\frac{1}{2}r^{2}\sinh \theta \partial _{r}+\frac{1}{2}r\cosh \theta
\partial _{\theta }  \label{CLN.150}\,.
\end{eqnarray}
We remind the reader that the variables $r$ and $\theta$ are 
defined in Eq.(\ref{CLN.29}).

Writing ~$L_{X^{I}}g_{ij}=2C_{I}\left( r,\theta \right) g_{ij}$ we find the
conformal factors of the CKVs $X^{I}$ $I=1,...6$ above in terms of the the
conformal function. The result is:%
%\begin{eqnarray}
%C_{1}\left( r,\theta \right) &=&-\frac{1}{r}\sinh \left( \theta \right) 
%\frac{N_{,\theta }}{N}~,~C_{2}\left( r,\theta \right) =-\frac{1}{r}\cosh
%\left( \theta \right) \frac{N_{,\theta }}{N}  \notag \\
%C_{3}\left( r,\theta \right) &=&\frac{N_{,\theta }}{N}~~,~C_{4}\left(
%r,\theta \right) =1,\;\;
%C_{5}\left( r,\theta \right) =\frac{r}{2}\left( \frac{%
%2N\cosh \theta +\sinh \theta N_{\theta }}{N}\right)  \notag \\
%C_{6}\left( r,\theta \right) &=&\frac{r}{2}\left( \frac{2N\sinh \theta
%+\cosh \theta N_{\theta }}{N}\right).  
%\label{CLN.153}
%\end{eqnarray}

\begin{equation*}
C_{1}\left( r,\theta \right)=-\frac{1}{r}\sinh\theta  
\frac{N_{,\theta }}{N},\;\;C_{2}\left( r,\theta \right) =-\frac{1}{r}\cosh\theta  \frac{N_{,\theta }}{N}  
\end{equation*}
\begin{equation*}
C_{3}\left( r,\theta \right)=\frac{N_{,\theta }}{N},\;\;C_{4}\left(
r,\theta \right) =1
\end{equation*}
\begin{equation*}
C_{5}\left( r,\theta \right) =\frac{r}{2}\left( \frac{%
2N\cosh \theta +\sinh \theta N_{\theta }}{N}\right)
\end{equation*}
\begin{equation*}
C_{6}\left( r,\theta \right)=\frac{r}{2}
\left( \frac{2N\sinh \theta+\cosh \theta N_{\theta }}{N}\right).  
%\label{CLN.153}
\end{equation*}

We would like to remind the reader that 
the coupling function $N(\theta)$ does not obey Eq.(\ref{CLN.300}),
otherwise the kinetic metric
of the Lagrangian (\ref{CLN.150a}) is flat ($R_{KM}$ vanishes) and
we return to Sec. 3. The latter means that the vectors $X^{I}$$
I=1,...6$, except the $I=4$, are proper CKVs therefore they do not give (if proper) a Noether point symmetry. The vector $X_{4}$ is a non-gradient HV which also
does not  produce a Noether point symmetry. Therefore, according to 
theorem in  \cite{TsamGE,TsampJP},  only Killing vectors are
possible to serve as Noether symmetries. Killing vectors 
do not exist in general but only for special
forms of the conformal function $N(\theta )$. Each of such forms of $N(\theta )$
results in a potential $V(\theta )\ $hence in  a scalar field potential which
admits Noether point symmetries. In the following,  we shall  determine the possible $%
N(\theta )$  forms which lead to a KV and give the corresponding Noether point symmetry
and the corresponding Noether integral which will be used for the solution
of the field equations.

\subsection{ Searching for Noether symmetries}
\begin{enumerate}
\item If $N\left( \theta \right) =\frac{N_{0}}{\cosh2\theta 
-1}$ then $X^{5}$ is a non-gradient KV and a Noether symmetry of the
Lagrangian (\ref{CLN.150a}) for the potential 
\begin{equation}
V\left( \theta \right) =\frac{V_{0}}{\cosh 2\theta  -1}~\text{%
or }V\left( \theta \right) =0 \,. \label{CLN.155}
\end{equation}%
The corresponding Noether integral is 
\begin{equation}
I_{X^{5}}=\frac{N_{0}^{2}r^{2}}{\left( \cosh2\theta 
-1\right) ^{2}}\left( r\dot{\theta}\sinh \theta -\dot{r}\cosh \theta \right)
.  \label{CLN.156}
\end{equation}

\item If $N\left( \theta \right) =\frac{N_{0}}{\cosh 2\theta 
+1}$ then $X^{6}$ is a non gradient KV, $X^{6}$ and a Noether symmetry for
the Lagrangian (\ref{CLN.150a}) if 
\begin{equation}
V\left( \theta \right) =\frac{V_{0}}{\cosh 2\theta  +1}~\text{%
or }V\left( \theta \right) =0  \label{CLN.157}\,.
\end{equation}%
The corresponding Noether integral is 
\begin{equation}
I_{X^{6}}=\frac{N_{0}^{2}r^{2}}{\left( \cosh2\theta 
+1\right) ^{2}}\left( r\dot{\theta}\cosh \theta -\dot{r}\sinh \theta \right)\,.
\label{CLN.158}
\end{equation}

\item If $N\left( \theta \right) =\frac{N_{0}}{\cosh ^{2}\left( \theta
+\theta _{0}\right) }$ then the linear combination $%
X^{56}=c_{1}X^{5}+c_{2}X^{6}$ where $c_{1}=\sinh \left( \theta _{0}\right) $
and $c_{2}=\cosh \left( \theta _{0}\right) $. $X^{56}$ is a Noether symmetry
for the Lagrangian (\ref{CLN.150a}) if 
\begin{equation}
V\left( \theta \right) =\frac{V_{0}}{\cosh ^{2}\left( \theta +\theta
_{0}\right) }~\text{or }V\left( \theta \right) =0  \label{CLN.159}
\end{equation}%
with Noether integral 
\begin{equation*}
I_{X^{56}}=\frac{N_{0}^{2}r^{2}}{\cosh ^{4}\left( \theta +\theta _{0}\right) 
}\left[ r\dot{\theta}\cosh \left( \theta +\theta _{0}\right) -\dot{r}\sinh
\left( \theta +\theta _{0}\right) \right]
\end{equation*}%
Obviously the third case is the most general situation 
and it contains cases 1 and 2  (and the
trivial case)\ as special cases. Therefore,  in the following,  we look for
analytic solutions for the vector $X^{56}$ only.

We recall that $\frac{1}{\sqrt{-2F\left( \theta \right) }}=N^{2}\left(
\theta \right) $ from which follows:%
\begin{equation}
~F\left( \theta \right) =-\frac{1}{2N_{0}^{4}}\cosh ^{8}\left( \theta
+\theta _{0}\right) ~,N_{0}\in 
%TCIMACRO{\U{211d} }%
%BeginExpansion
\mathbb{R}
%EndExpansion
.  \label{CLN.161}
\end{equation}
We may consider $\theta _{0}=0$ (e.g. by introducing the new variable $%
\Theta =\theta +\theta _{0}).$

For the potential (\ref{CLN.159}) Lagrangian (\ref{CLN.150a}) becomes 
\begin{equation}
L=\frac{N_{0}^{2}}{\cosh ^{4}\theta }\left( -\frac{1}{2}\dot{r}^{2}+\frac{1}{%
2}r^{2}\dot{\theta}^{2}\right) -r^{2}\frac{V_{0}}{\cosh ^{2}\theta }
\label{CLN.162}
\end{equation}%
and the Hamiltonian 
\begin{equation}
E=\frac{N_{0}^{2}}{\cosh ^{4}\theta }\left( -\frac{1}{2}\dot{r}^{2}+\frac{1}{%
2}r^{2}\dot{\theta}^{2}\right) +r^{2}\frac{V_{0}}{\cosh ^{2}\theta }.
\label{CLN.163}
\end{equation}

The Euler-Lagrange equations provide the equations of motion:%
\begin{equation}
\ddot{r}+r\dot{\theta}^{2}-4\tanh \theta ~\dot{r}\dot{\theta}-2\frac{V_{0}}{%
N_{0}^{2}}r\cosh ^{2}\theta=0  \label{CLN.164} 
\end{equation}
\begin{equation}
\ddot{\theta}-2\tanh \theta ~\left( \frac{1}{r^{2}}\dot{r}^{2}+\dot{\theta}%
^{2}\right) +\frac{2}{r}\dot{r}\dot{\theta}-2\frac{V_{0}}{N_{0}^{2}}\cosh
\theta \sinh \theta =0  \label{CLN.165}
\end{equation}%
and the Noether integral $I$ for $\theta _{0}=0$ becomes: 
\begin{equation}
I=\frac{N_{0}^{2}r^{2}}{\cosh ^{4}\left( \theta +\theta _{0}\right) 
}\left[ r\dot{\theta}\cosh \theta  -\dot{r}\sinh \theta
 \right] .  \label{CLN.166}
\end{equation}

In order to proceed with the solution of the system of equations 
(\ref{CLN.164}), (\ref{CLN.165}) we change to the coordinates $x,y~$which we
define by the relations%
\begin{equation}
r=\frac{x}{\sqrt{1-x^{2}y^{2}}}~,~\theta =\arctan h\left( xy\right) .
\label{CLN.168}
\end{equation}%
In the coordinates $(x,y)$ the Lagrangian and the Hamiltonian are written  as
\begin{equation}
L=\frac{N_{0}^{2}}{2}\left( -\dot{x}^{2}+x^{4}\dot{y}\right) -V_{0}x^{2}
\label{CLN.169}
\end{equation}%
\begin{equation}
E=\frac{N_{0}^{2}}{2}\left( -\dot{x}^{2}+x^{4}\dot{y}^{2}\right) +V_{0}x^{2}
\label{CLN.170}
\end{equation}%
and the Noether integral is %(we write $I$ for $I_{X^{56}}!)$%
\begin{equation}
%I=N_{0}^{2}x^{4}\dot{y}.  
I=x^{4}\dot{y}.  
\label{CLN.171}
\end{equation}%
%Let us assume that $I\neq 0.$

In the new variables,  the Euler-Lagrange equations read:%
\begin{eqnarray}
\ddot{x}+2x^{3}\dot{y}^{2}-\frac{2V_{0}}{N_{0}^{2}}x &=&0  \label{CLN.172} \\
\ddot{y}+\frac{4}{x}\dot{x}\dot{y} &=&0.  \label{CLN.173}
\end{eqnarray}%
In this context, from the Noether integral,  we have%
\begin{equation}
\dot{y}=\frac{I}{x^{4}}  \label{CLN.174}
\end{equation}%
which, upon substitution in the field equations, gives the system:%
\begin{eqnarray}
\ddot{x}+\frac{2I^{2}}{x^{5}} -\frac{2V_{0}}{N_{0}^{2}}x&=&0  \label{CLN.175}
\\
\frac{N_{0}^{2}}{2}\left( -\dot{x}^{2}+\frac{I^{2}}{x^{4}}\right)
+V_{0}x^{2} &=&E.  \label{CLN.176}
\end{eqnarray}%
from which we compute%
\begin{equation}
\dot{x}=\sqrt{\frac{I^{2}}{x^{4}}+\frac{2V_{0}}{N_{0}^{2}}x^{2}-\frac{2E}{%
N_{0}^{2}}}  \label{CLN.179}
\end{equation}%
and the analytical solution 
\begin{equation}
\int \frac{dx}{\sqrt{\frac{I^{2}}{x^{4}}+\frac{2V_{0}}{N_{0}^{2}}x^{2}-\frac{%
2E}{N_{0}^{2}}}}=t-t_{0}.  
\label{CLN.180}
\end{equation}%
Also, integrating Eq.(\ref{CLN.174}), we find 
\begin{equation}
y\left( t\right) -y_{0}=\int \frac{I}{x^{4}}dt.  \label{CLN.181}
\end{equation}

If we consider the special case where $I=0$ then the analytic solution is%
\begin{equation}
x=x_{0}\sinh \left( \frac{\sqrt{2V_{0}}}{N_{0}}t+x_{1}\right) ,~y=y_{0}
\end{equation}%
with the Hamiltonian constrain $E=-x_{0}^{2}V_{0}.$

\item Finally, if $V_{0}=0~$(i.e. free particle) and 
$I=0$ the analytic solution becomes 
\begin{equation}
x=x_{0}t+x_{1}~,~y=y_{0}
\end{equation}%
with Hamiltonian constrain $E=-\frac{1}{2}x_{0}^{2}N_{0}^{2}$.

\end{enumerate}

\section{Conclusions}
In this work we have identified the Noether point symmetries of the equations
of motion in the context of scalar-tensor cosmology considering a 2-dimensional minisuperspace ${\cal Q}=\{\psi, a\}$. 
We find that there is a rather large class of hyperbolic and exponential 
potentials which admit extra (beyond the $\partial_{t}=0$) 
Noether symmetries which lead to 
integral of motions.
This approach is extremely efficient in physical problems since 
it can be utilized in order to simplify a given system of differential
equations as well as to determine the integrability of the system. 
Based on the above arguments, we manage to provide 
general 
analytical solutions in scalar-tensor cosmologies assuming a FRW spatially flat metric.
These solutions can be used in order to 
compare cosmographic parameters, 
such as  the  Hubble expansion
rate, the deceleration parameter, snap, jerk and density parameters with 
observations \cite{bamba}. Such an analysis is in progress and it will 
be published in a forthcoming paper.

\section*{Acknowledgments}
SB acknowledges support by the Research
Center for Astronomy of the Academy of Athens in the context of the
program {\it ``Tracing the Cosmic Acceleration''}. 

\appendix{\section{Maximally symetric space: the case  $|k|=1$}}
In order to complete Sec.3,  we provide here 
the main steps of the Noether algebra in the case where $k=\pm 1$.
Briefly, we start with the KVs of the kinematic metric  
\begin{equation*}
K^{1}=\frac{1}{N_{0}^{2}}\frac{e^{-2k\theta }}{r}\left(k \partial _{r}+\frac{1%
}{r}\partial _{\theta }\right),~K^{2}=\frac{1}{N_{0}^{2}}\left(
-kr\partial _{r}+\partial _{\theta }\right), 
\end{equation*}
\begin{equation*}
K^{3}=-r\left[ \ln \left(
re^{-k\theta }\right) -1\right] \partial _{r}+\ln \left( re^{-k\theta }\right)
\partial _{\theta }  
\end{equation*}
where the vectors $K^{1,2}$ are gradient and $K^{3}$ is non-gradient. 
Also the HV is given by
\begin{equation*}
H^{i}=\frac{1}{4}r\left[ 2\ln \left( re^{-k\theta }\right) +3\right] \partial
_{r}-\frac{1}{2}\left[ \ln \left( re^{-k\theta }\right) 
+\frac{1}{2}\right]
\partial _{\theta }  \;.
\end{equation*}

Using the theorem in  \cite{TsamGE,TsampJP} 
and making some simple calculations (see Sec. 3) 
we find the following results:

\begin{enumerate}
\item Noether symmetries generated by the KV $K^{1}$.

a) If $V\left( \theta \right) =V_{0}e^{-2k\theta }$ then we have the Noether
symmetries $K^{1}~,~tK^{1}$ with Noether integrals%
\begin{equation*}
I_{1}^{\prime }=\frac{d}{dt}\left(k \theta -\ln r\right),\;I_{2}^{\prime }=t%
\left[ \frac{d}{dt}\left(k\theta -\ln r\right) \right] -\left(k \theta -\ln
r\right)  
\end{equation*}

b) If $V\left( \theta \right) =V_{0}e^{-2k\theta }-\frac{1}{4}pe^{2k\theta }\,$
then we have the Noether symmetries $K^{1}~,~tK^{1}$ with Noether integrals%
\begin{equation*}
I_{1}=\frac{d}{dt}\left(k \theta -\ln r\right)-pt
\end{equation*}
\begin{equation*}
I_{2}=t\left[ \frac{d}{dt}\left(k \theta -\ln r\right) \right] -\left(k \theta
-\ln r\right) -\frac{1}{2}pt^{2}  
\end{equation*}

\item Noether symmetries generated by the KV $K^{2}$.

a) If $V\left( \theta \right) =V_{0}e^{2k\theta }~$then we have the extra
Noether symmetries $K^{2}~,~tK^{2}$ with Noether integrals 
\begin{equation*}
J_{1}=\left[ \frac{d}{dt}\left( \frac{1}{2}e^{2k\theta }r^{2}\right) \right],~J_{2}=t\left[ \frac{d}{dt}\left( \frac{1}{2}e^{2k\theta }r^{2}\right) %
\right] -\frac{1}{2}e^{2k\theta }r^{2}  \label{CLN.74}
\end{equation*}%
b) If $V\left( \theta \right) =\left( V_{0}e^{2k\theta }-\frac{m}{2}k\theta
e^{2k\theta }\right) $, then we have the Noether 
symmetries $e^{\pm \sqrt{m}%
t}K^{2}$ with Noether integrals%
\begin{equation*}
J_{1,2}^{\prime }=e^{\pm \sqrt{m}t}\left( \left[ \frac{d}{dt}\left( \frac{1}{%
2}e^{2\theta }r^{2}\right) \right] \mp \frac{\sqrt{m}}{2}e^{2\theta
}r^{2}\right)  
\end{equation*}

\item If $V\left( \theta \right) =0$ then 
the system becomes the free particle (see \cite{TsamGE}).
\end{enumerate}
To this end the corresponding analytical solutions can be found utilizing the 
above integrals the arguments of Sec. 3 and the new coordinates 
$(u,v)=(k\theta-{\rm ln}r,\frac{1}{2}{\rm e}^{2k\theta}r^{2})$. 

%%%%%%%%%%%%%%%%%%%%%%%%%%%%%%%%%%%%%%%%%%%%%%%%%%%%%%%%%%%%%%%%%%%%%%%%%
%%%%%%%%%%%%%%%%%%%%%%%%%%%%%%%%%%%%%%%%%%%%%%%%%%%%%%%%%%%%%%%%%%%%%%%%%

\end{document}